\documentclass[aps,prl,twocolumn,superscriptaddress,floatfix,longbibliography,10pt]{revtex4-2}

\usepackage{siunitx}
\DeclareSIUnit{\electronvolt}{eV}

\usepackage{stix} 

\usepackage[english]{babel}
\usepackage{graphicx}
\usepackage{hyperref}
\usepackage{xcolor}
\usepackage{amsmath}
\usepackage[T1]{fontenc}
\usepackage{textcomp,gensymb}
\usepackage{pifont} 
\usepackage{orcidlink} 

\renewcommand{\selectlanguage}[1]{}

\makeatletter
\@ifclasswith{revtex4-2}{preprint}
  {\newif\ifpreprint\preprinttrue}
  {\newif\ifpreprint\preprintfalse}
\makeatother

\ifpreprint
    \usepackage{lineno}
    \linenumbers
\fi

\begin{document}

\title{Dressed-State Spectroscopy of Proton Spins in Water Beyond the Rotating-Wave Approximation}

\author{Ivo~Schulthess\orcidlink{0000-0002-5621-2462}}
\email[Corresponding author: ]{ivo.schulthess@ethz.ch}
\affiliation{Laboratory for High Energy Physics and Albert Einstein Center for Fundamental Physics, University of Bern, 3012 Bern, Switzerland}
\affiliation{Institute for Particle Physics and Astrophysics, ETH Zurich, 8093 Zurich, Switzerland}

\author{Anastasio~Fratangelo\orcidlink{0000-0001-9964-601X}}
\altaffiliation[Current address: ]{Los Alamos National Laboratory, Los Alamos, New Mexico, 87545, USA}
\affiliation{Laboratory for High Energy Physics and Albert Einstein Center for Fundamental Physics, University of Bern, 3012 Bern, Switzerland}

\author{Patrick~Hautle\orcidlink{0000-0002-0502-8278}}
\affiliation{Laboratory for Neutron and Muon Instrumentation (LIN), Paul Scherrer Institut, 5232 Villigen PSI, Switzerland}

\author{Philipp~Heil\orcidlink{0000-0001-5309-5988}}
\affiliation{Laboratory for High Energy Physics and Albert Einstein Center for Fundamental Physics, University of Bern, 3012 Bern, Switzerland}

\author{Gjon~Markaj}
\affiliation{Laboratory for High Energy Physics and Albert Einstein Center for Fundamental Physics, University of Bern, 3012 Bern, Switzerland}

\author{Marc~Persoz\orcidlink{0009-0002-6748-6837}}
\affiliation{Laboratory for High Energy Physics and Albert Einstein Center for Fundamental Physics, University of Bern, 3012 Bern, Switzerland}

\author{Ciro~Pistillo\orcidlink{0000-0001-8131-9440}}
\affiliation{Laboratory for High Energy Physics and Albert Einstein Center for Fundamental Physics, University of Bern, 3012 Bern, Switzerland}

\author{Jacob~Thorne\orcidlink{0000-0002-3905-5549}}
\affiliation{Laboratory for High Energy Physics and Albert Einstein Center for Fundamental Physics, University of Bern, 3012 Bern, Switzerland}

\author{Florian~M.~Piegsa\orcidlink{0000-0002-4393-1054}}
\affiliation{Laboratory for High Energy Physics and Albert Einstein Center for Fundamental Physics, University of Bern, 3012 Bern, Switzerland}

\date{\today}

\begin{abstract}
    The quantum Rabi model provides the framework for describing a two-level system interacting with a strong oscillating field beyond the rotating-wave approximation. We report the first experimental observation of the resulting dressed states of proton spins in water, realized using a Rabi-type setup with a strong off-resonant magnetic dressing field. The measured resonance spectrum exhibits multiple spin-state transitions involving several dressing-field quanta, including higher-order resonances predicted by the quantum Rabi model. The dressed-state energies show excellent agreement with theoretical expectations, extending dressed-state spectroscopy to proton spins and opening new possibilities for precision spin manipulation in nuclear magnetic resonance and related precision measurements.
\end{abstract}

\maketitle

\ifpreprint
    \section{Introduction}
\fi

Dressed states emerge when a quantum system is driven by an additional off-resonant oscillating field that modifies the energy levels and gives rise to multi-photon transitions~\cite{Autler:1955zz, cohen-tannoudji_absorption_1969}. Such interactions are described by the quantum Rabi model. Dressing effects of nuclear systems have been investigated with neutrons~\cite{Golub:1994cg, muskat_dressed_1987, Dubbers:1989pa} and $^3$He~\cite{Esler:2007dt, Chu:2010me}, and with various atomic systems~\cite{haroche_modified_1970, yabuzaki_modification_1972, bartenev_anomalous_1975, Zanon-Willette:2012lhm, arias_realization_2019}, typically using weak to moderate dressing-field amplitudes and observed as shifts of the main resonance frequency. Dressed states also play a central role in quantum optics and quantum information, where both weak and strong driving regimes have been extensively explored, including the emergence of multi-photon resonances and strongly modified eigenstates~\cite{cohentannoudji_atomphoton_1998, Haroche:2006bmm, Blais:2020wjs}. Such systems are increasingly exploited as a resource for quantum computation and quantum control, for example in dressed‑spin qubit architectures in semiconductors~\cite{Laucht:2016jiq,Seedhouse:2021lpu}. In nuclear spin systems, strong driving of proton spins has been demonstrated in nitrogen-vacancy-based nuclear magnetic resonance (NMR), but without resolving a dressed-state spectrum or higher-order multi-photon resonances~\cite{Yudilevich:2023uev}. In contrast, in this work we study the dressed states of proton spins in water using an off-resonant dressing field in a Rabi-type apparatus and directly map the associated multi-photon dressed-state transitions beyond the rotating-wave approximation.

In our system, the nuclear spin of the proton in water serves as a two-level system subject to a static magnetic field $B_0$ and an off-resonant oscillating dressing field $B_d \cos(\omega_d t)$.\footnote{The only relevant difference to a free proton is a chemical shift of the Larmor frequency, which is negligible compared to the experimental linewidth and does not affect the dressed-state structure discussed here.} The dressing field modifies the effective energy splitting and generates a non-trivial dressed-state structure, including transitions involving multiple photons. For weak, near-resonant driving (i.e.\ $\gamma B_d \ll \omega_0$, $\gamma B_d \ll \omega_d$, and $|\omega_0-\omega_d| \ll \omega_0$, where $\gamma$ is the proton gyromagnetic ratio), the rotating-wave approximation applies, under which the quantum Rabi model reduces to the Jaynes--Cummings model~\cite{Jaynes:1963zz}. However, in the parameter regime explored here, characterized by significant detuning and coupling strength comparable to the Larmor frequency ($\gamma B_d \sim \omega_0$), the counter-rotating contribution becomes significant. This contribution arises from the component of the linearly oscillating field that rotates opposite to the spin precession and is neglected under the rotating-wave approximation. In this regime, the quantum Rabi model provides the appropriate description, capturing both the energy-level shifts and the emergence of higher-order transitions.

Previous nuclear spin-dressing studies operated in the weak-driving regime, and only shifts of the main resonance were probed. In particular, in the high-frequency dressing limit, the observed behavior followed the characteristic Bessel-function dependence obtained within the rotating-wave approximation, reflecting the dominance of the co-rotating component of the dressing field and the suppression of counter-rotating effects. In contrast, the present work operates with an off-resonant dressing field and amplitudes large enough that counter-rotating contributions significantly modify the energy structure. Under these conditions, a ladder of avoided crossings and multi-photon transitions predicted by the quantum Rabi model emerges, which we observe here for the first time in a nuclear spin system in quantitative agreement with the model. Mapping this spectrum provides a stringent test of the theory and accesses dressed-state physics far beyond the small-amplitude limit. Here, the counter-rotating component of the drive field becomes relevant for precision spin-manipulation experiments, producing the Bloch--Siegert shift~\cite{Bloch:1940jgs, ramsey_resonance_1955}. Off-resonant dressing fields have been proposed as a means of compensating this shift~\cite{greene_observation_1978}.

\ifpreprint
    \section{Setup}
\fi

\begin{figure}[!tb]
    \centering
    \ifpreprint
        \includegraphics[width=0.96\textwidth]{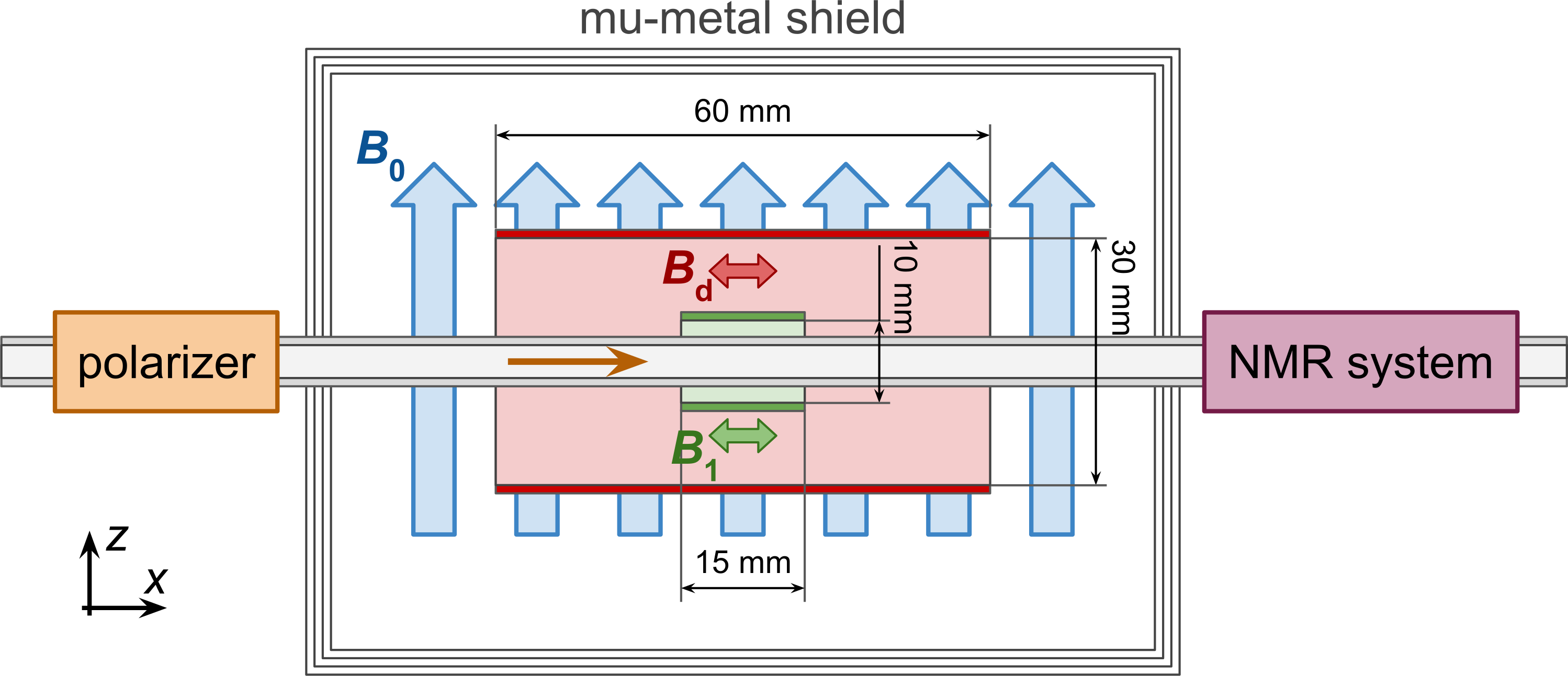}
    \else
        \includegraphics[width=0.48\textwidth]{figures/schematics.png}
    \fi
    \caption{Schematic of the experimental setup (not to scale). The protons in flowing water enter the setup from the left. After polarization (orange), they continue into the interaction region, which is shielded by a four-layer mu-metal shield (gray). The innermost layer has a diameter of \SI{180}{\milli\meter} and a length of \SI{360}{\milli\meter}. The constant main magnetic field $B_0$ (blue) is in the $z$-direction. The oscillating fields for the spin dressing $B_d$ (red) and the spin flip $B_1$ (green) are oriented in the water flow direction along the $x$-axis and can be applied via their corresponding coils. A pulse-NMR system (purple) is used to measure the spin polarization. }
    \label{fig:schematic}
\end{figure}

We measured the dressed proton spin states using the apparatus described in~\cite{Schulthess:2023mgx}. In the present Rabi-type configuration, the oscillating dressing field ${B_d \cos(\omega_d t)}$ and the spin-flip field ${B_1 \cos(\omega_1 t)}$ are generated by two coaxial solenoid coils oriented along the water-flow direction, i.e.\ orthogonal to $B_0$. The larger dressing-field coil with a diameter of \SI{30}{\milli\meter} and 69~windings over a length of \SI{60}{\milli\meter} surrounds the spin-flip coil which has a diameter of \SI{10}{\milli\meter}, a length of \SI{15}{\milli\meter}, and 16~windings. In principle, the dressing field could also be applied via the spin-flip coil but a dedicated coil can reduce fringe-field effects. They are positioned in the center of a cylindrical four-layer magnetic mu-metal shield with endcaps on each layer, providing a shielding factor exceeding $10^6$~\cite{twinleaf_llc_twinleaf_2022}. The static main magnetic field $B_0 \approx \SI{23.5}{\micro\tesla}$ is created by the internal coils of the shield. A schematic of the setup is shown in Fig.~\ref{fig:schematic}.
The proton spins of hydrogen in water serve as the two-level system. The water is circulated through the apparatus by a gear pump. For the measurements reported here, we measured a flow velocity of \SI{1.60(3)}{\meter\per\second}. The water temperature was stabilized and remained at \SI{20.60(4)}{\celsius} during operation, inferred from the temperature measured at the water-cooled NMR magnet. The proton spins are first polarized in a static magnetic field of approximately \SI{190}{\milli\tesla}, resulting in a thermal polarization of the order $P_0 \approx 7 \times 10^{-7}$. The water then flows through a glass capillary with an inner diameter of \SI{3}{\milli\meter} to the interaction region. The resulting spin polarization is measured using a commercial pulse-NMR system, operating at about \SI{0.5}{\tesla} or \SI{22}{\mega\hertz}~\cite{spincore_technologies_inc_ispin-nmr_2017}.

\ifpreprint
    \section{Formalism}
\fi

The spin--field interaction is described by the quantum Rabi Hamiltonian for a two-level system coupled to a single mode of the dressing field, 
\begin{equation}\label{eq:QRM}
    \mathcal{\hat{H}} = \underbrace{\vphantom{\left(a^\dag\right)} \omega_0 \hat{s}_z + \hbar \omega_d \hat{a}^\dagger \hat{a}}_{\mathcal{\hat{H}}_0} + \underbrace{\lambda \hat{s}_x \left( \hat{a} + \hat{a}^\dagger \right)}_{\mathcal{\hat{V}}_\mathrm{int}} \, , 
\end{equation}
which, in our case, models the proton spin interacting with the transverse oscillating magnetic field. The spin-flip probe field $B_1$ is treated separately as a weak perturbation used to induce transitions between dressed states. The first term $\mathcal{\hat{H}}_0$ describes the uncoupled system: the proton spin with Larmor frequency $\omega_0$, where $\hat{s}_z = \frac{\hbar}{2} \hat{\sigma}_z$, and the dressing-field mode of frequency $\omega_d$, with $\hat{a}^\dagger$ and $\hat{a}$ denoting photon creation and annihilation operators, respectively. The interaction term $\mathcal{\hat{V}}_\mathrm{int}$ involves the transverse spin operator $\hat{s}_x = \frac{\hbar}{2} \hat{\sigma}_x$ and a coupling strength 
\begin{equation}\label{eq:coupling}
    \lambda = \frac{\gamma B_d}{2 \sqrt{\bar{n}}} \ ,
\end{equation}
where $\bar{n} \gg 1$ is the average photon number in the dressing field. By introducing the dimensionless dressing parameters 
\begin{equation}\label{eq:dressingParameters}
    x \equiv \frac{\gamma B_d}{\omega_d} \, , \quad y \equiv \frac{\gamma B_0}{\omega_d} = \frac{\omega_0}{\omega_d} \, ,
\end{equation}
the Hamiltonian Eq.~(\ref{eq:QRM}) can be rewritten as
\begin{equation}\label{eq:QRM_dressing}
    \frac{\mathcal{\hat{H}}}{\hbar \omega_d} = \frac{y}{2} \hat{\sigma}_z + \hat{a}^\dagger \hat{a} + \frac{x}{4 \sqrt{\bar{n}}} \hat{\sigma}_x \left( \hat{a} + \hat{a}^\dagger \right) \, .
\end{equation}
Although the dressing field in our experiment contains a large number of photons ($\bar{n} \gg 1$) and thus behaves classically, the quantum Rabi model provides a compact description of the spin-field interaction. This framework captures the full structure of the dressed-state spectrum, including counter-rotating contributions and multi-photon processes, which are neglected under the rotating-wave approximation. 

To obtain a matrix representation, we can expand the Hamiltonian of Eq.~(\ref{eq:QRM_dressing}) in the Fock basis~\cite{fock_konfigurationsraum_1932}, ordered as $(...,~|g,n{-}1\rangle,~|e,n\rangle,~|g,n\rangle,~|e,n{+}1\rangle,~...)$ , where $g$ and $e$ denote the ground and excited spin states, respectively and $n$ is the photon number of the dressing-field mode. This leads to the block-tridiagonal matrix 
\begin{widetext}
    \begin{equation}\label{eq:QRM_matrix}
        \mathcal{\hat{H}} = \left(\begin{smallmatrix}
        \ddots & \vdots & \vdots & \vdots & \vdots & \vdots & \vdots & \adots \\
        \cdots & \phantom{-}\frac{y}{2}+(n+1) & 0 & 0 & x/4 & 0 & 0 & \cdots \\
        \cdots & 0 & -\frac{y}{2}+(n+1) & \frac{x}{4} & 0 & 0 & 0 & \cdots \\
        \cdots & 0 & \frac{x}{4} & \phantom{-}\frac{y}{2}+n\phantom{(-1)} & 0 & 0 & x/4 & \cdots \\
        \cdots & x/4 & 0 & 0 & -\frac{y}{2}+n\phantom{(-1)} & \frac{x}{4} & 0 & \cdots \\
        \cdots & 0 & 0 & 0 & \frac{x}{4} & \phantom{-}\frac{y}{2}+(n-1) & 0 &  \cdots\\
        \cdots & 0 & 0 & x/4 & 0 & 0 & -\frac{y}{2}+(n-1) & \cdots \\
        \adots & \vdots & \vdots & \vdots & \vdots & \vdots & \vdots & \ddots \\
        \end{smallmatrix}\right) \, .
    \end{equation}
\end{widetext}
The diagonal elements arise from the uncoupled Hamiltonian $\mathcal{\hat{H}}_0$, while the off-diagonal elements originate from the interaction term $\mathcal{\hat{V}}_\mathrm{int}$ and couple spin flips to photon creation and annihilation processes in the dressing-field mode. Unlike the Jaynes--Cummings model, obtained under the rotating-wave approximation, the quantum Rabi model does not conserve the total excitation number~\cite{Larson:2021flt}. This is reflected in the presence of the outer off-diagonal coupling terms (e.g.\ the $x/4$ matrix elements connecting $|g,n\rangle$ and $|e,n{+}1\rangle$), which arise from the counter-rotating contribution of the interaction. They are highlighted in Eq.~(\ref{eq:QRM_matrix}) using the slash-fraction notation. Under the rotating-wave approximation these matrix elements are absent, resulting in a block-diagonal structure. In the high-photon-number regime probed here, the non-conservation of excitation number manifests in observable features such as Bloch--Siegert shifts and the emergence of multiple resonances for a single dressing-field amplitude. These effects cannot be captured within the rotating-wave approximation. The matrix representation thus highlights the key physical differences between the two descriptions and motivates the use of the quantum Rabi framework for interpreting our measurements.

\begin{figure}[!tb]
    \centering
    \ifpreprint
        \includegraphics[width=0.96\textwidth]{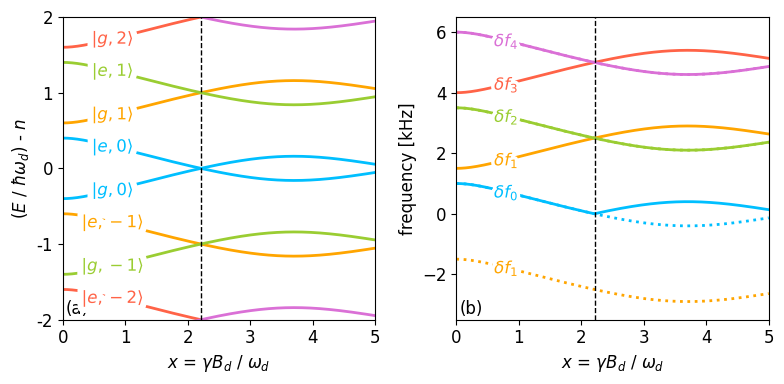}
    \else
        \includegraphics[width=0.48\textwidth]{figures/energyLevels.png}
    \fi
    \caption{(a) Energy level diagram, calculated from Eq.~(\ref{eq:QRM_matrix}). It shows the normalized dressed state energy as a function of the dressing parameter $x$ for $y=0.8$. The labels indicate the corresponding Fock state at $x=0$. The vertical dashed line indicates the position of a level crossing at $x\approx2.2$. (b) Calculated transition frequencies for the energy levels shown. Since negative frequencies (dotted levels) cannot be distinguished in our measurements, they are mapped to positive values. }
    \label{fig:energyLevels}
\end{figure}

The Hamiltonian matrix in Eq.~(\ref{eq:QRM_matrix}) can be numerically diagonalized for any choice of dressing parameters. The resulting eigenvalues correspond to the dressed-state energies, while the eigenvectors $|\psi_i\rangle$ are superpositions of the Fock-basis states. An example calculation for $\omega_0 = 2 \pi \times \SI{1000}{\hertz}$ and $\omega_d = 2 \pi \times \SI{1250}{\hertz}$, resulting in $y = \omega_0/\omega_d = 0.8$, is shown in Fig.~\ref{fig:energyLevels}. This choice of $\omega_d > \omega_0$ connects to the parameter regime commonly explored in earlier nuclear spin dressing experiments~\cite{muskat_dressed_1987, Esler:2007dt, Chu:2010me}. It leads to a reduction of the main resonance $\delta f_0$, which approximately follows a Bessel-function behavior expected from the rotating-wave approximation, represented by the blue line in Fig.~\ref{fig:energyLevels}(b). A matrix dimension of $50 \times 50$, including up to ${n \pm 12}$ photon states, was used. For the largest dressing parameters shown, the resulting transition frequencies deviate from fully converged values at the 1-Hz level, well below the experimental linewidth. 
Figures~\ref{fig:energyLevels}(a) and \ref{fig:energyLevels}(b) show the corresponding dressed-state energies and transition frequencies, respectively. Higher-order transitions $\delta f_n$ arise from effective multi-photon processes involving the exchange of more than one dressing-field quanta. The off-diagonal interaction terms in Eq.~(\ref{eq:QRM_matrix}) generate the characteristic level repulsion. For example, the $|g,0\rangle$ and $|e,1\rangle$ states undergo an avoided crossing near $x \approx 3.7$. In contrast, states such as $|g,0\rangle$ and $|e,0\rangle$ do not couple and therefore cross at e.g. $x \approx 2.2$.

For each transition, the matrix element is $m_{ij} = \langle \psi_i | \hat{V}_1 | \psi_j \rangle$, where $\hat{V}_1$ represents the coupling to the spin-flip field $B_1$. In the Fock basis it takes the form $\hat{V}_1 = \mathbb{I}_{N/2} \otimes \hat{\sigma}_x$, or explicitly
\begin{equation}\label{eq:probeMatrix}
    \hat{V}_1 = 
        \begin{pmatrix}
            \hat{\sigma}_x & 0 & \cdots & 0 \\
            0 & \hat{\sigma}_x & \cdots & 0 \\
            \vdots & \vdots & \ddots & \vdots \\
            0 & 0 & \cdots & \hat{\sigma}_x
        \end{pmatrix} \, . 
\end{equation}
Since the probe operator $\hat{V}_1$ flips only the spin degree of freedom without changing the photon number, transitions are allowed only between dressed states that share photon-number components with opposite spin character. The corresponding transition probability is $P_{ij} = | m_{ij} |^2$. Due to phase averaging over the continuous water flow which randomizes the spin phase relative to the dressing field, the measured resonance amplitudes reflect these transition probabilities.

\ifpreprint
    \section{Measurements}
\fi

\begin{figure*}[!tb]
    \centering
    \includegraphics[width=1.0\textwidth]{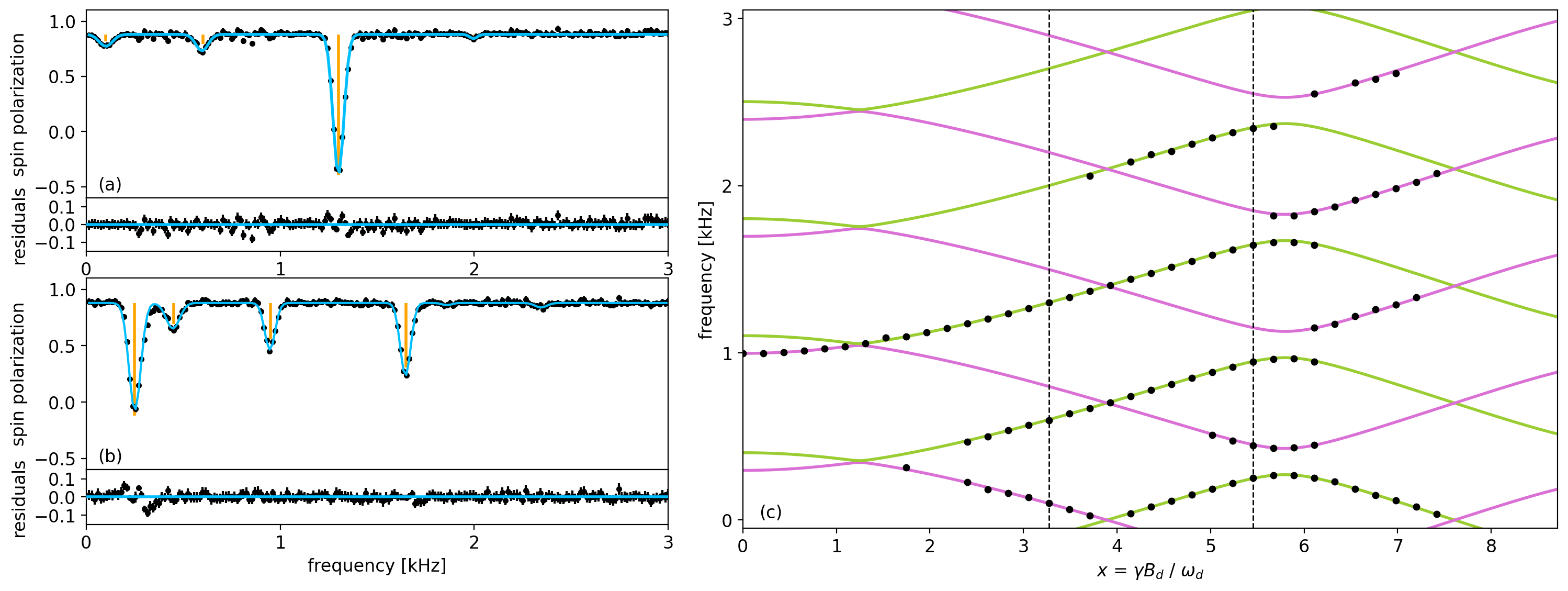}
    \caption{Measurements for the configuration $\omega_0 = 2 \pi \times 1000~\mathrm{Hz}$ and $\omega_d = 2 \pi \times 350~\mathrm{Hz}$, corresponding to $y\approx2.86$. (a) and (b) Two measured spectra (black dots) at $x=3.3$ ($B_d = \SI{26.9}{\micro\tesla}$) and $x=5.5$ ($B_d = \SI{44.8}{\micro\tesla}$), fitted with the multi-Gaussian function given in Eq.~(\ref{eq:multiGaussian}) (blue lines). The residuals are shown below each spectrum and confirm the quality of the fit used to extract the resonance frequencies. The scaled transition probabilities are shown at the corresponding calculated transition frequencies (orange vertical bars). (c) Energy level diagram showing the transition frequencies as a function of the dressing parameter $x$. The calculated energies are shown together with the fitted resonant transition frequencies (black dots). Error bars are smaller than the marker size. The purple and green lines distinguish the two families of transition branches and highlight (avoided) level crossings. The vertical dashed lines indicate the positions of the spectra shown in (a) and (b). }
    \label{fig:spectra}
\end{figure*}

Having established the theoretical framework, we now turn to the experimental investigation of the dressed-spin states for various combinations of $\omega_0$ and $\omega_d$. Here, we show only the measurements for $\omega_0 = 2 \pi \times 1000~\mathrm{Hz}$ ($B_0 \approx \SI{23.5}{\micro\tesla}$) and $\omega_d = 2 \pi \times 350~\mathrm{Hz}$, corresponding to $y\approx2.86$. The results of other configurations are available in the online repository~\cite{schulthess_ivoschulthessprotonnmr_dressedstates_2023}. The spectra were obtained by performing Rabi measurements with the spin-flip coil. To measure both, the main resonance and higher-order transitions, we scanned a broad frequency range between \SI{15}{\hertz} and \SI{3000}{\hertz} in steps of \SI{15}{\hertz}. We measured the spectra for a set of 41 dressing-field amplitudes, ranging from $B_d = \SI{0}{\micro\tesla}$ up to $B_d \approx \SI{183}{\micro\tesla}$. These ranges provide sufficient resolution to resolve all avoided crossings and higher-order resonances that remain clearly observable above the noise level. The resonant transition frequencies were extracted and compared with the dressed-state energies calculated. Because the dressing-field amplitude cannot be measured directly due to technical constraints, it was determined by fitting the calculated energy levels to the observed spectra, with $B_d$ as the only free parameter. Additionally, the calibration was verified through an offline measurement using constant currents applied to the dressing coil, with the resulting magnetic field measured by a Hall probe. Two spectra are shown in Figs.~\ref{fig:spectra}(a) and~\ref{fig:spectra}(b) together with the fits of a multi-Gaussian function 
\begin{equation}\label{eq:multiGaussian}
    G(f) = \sum_{i=1}^N \left( a_i \, \exp{\left( \frac{-\left( f - {f_0}_i\right)^2}{2 c_i^2} \right)} \right) + o \, ,
\end{equation}
where $N$ is the number of Gaussians fitted, $a_i$ their amplitudes, ${f_0}_i$ their center positions, $c_i$ their widths, and $o$ is a common offset. The Gaussian form provides an empirical description of the inhomogeneously broadened lineshapes. The fitted widths $c_i$ are distributed around \SI{27(4)}{\hertz}, where the quoted uncertainty reflects the spread of the fit results. This broadening is dominated by the finite length and fringe field of the spin-flip coil. The residuals, representing the difference between the measured data and the fit, are shown below each spectrum. The fits yield reduced $\chi^2$ of about 0.63 for the spectra shown, indicating that the multi-Gaussian model describes the data within their uncertainties.

Only transitions originating from populated dressed eigenstates contribute to the measured resonances. The initially prepared polarized spin state decomposes into a specific superposition of dressed eigenstates, which determines the observable transitions. The calculated transition probabilities $P_{ij}$, which quantify the intrinsic coupling strengths between dressed states, are shown as vertical lines in Figs.~\ref{fig:spectra}(a) and~\ref{fig:spectra}(b). To enable a comparison with the measured spectra, a global scaling factor was applied. This factor maps the dimensionless transition probabilities onto the experimentally observed polarization and accounts for various experimental effects, including incomplete polarization, depolarization, finite interaction times, and detection efficiency. While this procedure reproduces the relative strengths of the dominant resonances, deviations in the absolute amplitudes of individual lines remain, reflecting additional experimental effects not captured by the simplified transition-probability model. Figure~\ref{fig:spectra}(c) shows the full diagram with the calculated energy levels and the fitted resonance frequencies of all spectra. Only data points with $a_i / \sigma_{a_i} \geq 2$ are shown, where $\sigma_{a_i}$ is the uncertainty of the fitted amplitude. 

\begin{figure}[!tb]
    \centering
    \ifpreprint
        \includegraphics[width=0.96\textwidth]{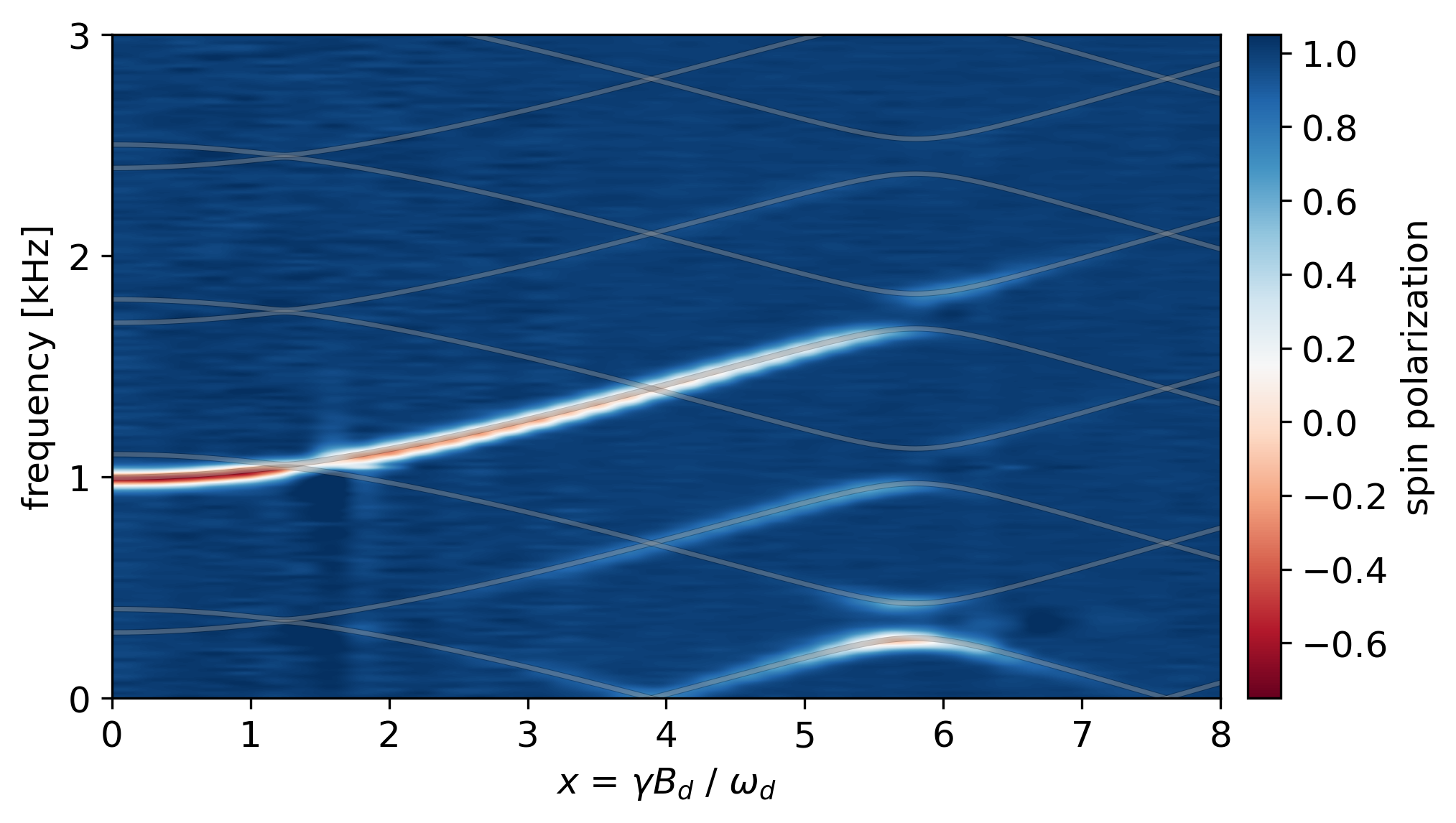}
    \else
        \includegraphics[width=0.48\textwidth]{figures/densityPlot.png}
    \fi
    \caption{Density map of the dressed spin states. It shows the spin polarization (color scale) as a function of the probe frequency and the dressing parameter $x$. This representation corresponds to the raw spectra as shown in the two examples in Fig.~\ref{fig:spectra} with only a baseline correction applied. The calculated energy levels are overlaid as semi-transparent solid lines to guide the eye. }
    \label{fig:densityPlot}
\end{figure}

In addition to fitting the spectra with Gaussian functions to find the resonance frequencies and thus the transition energies, complementary visual information is obtained by plotting the raw data in a two-dimensional density map as shown in Fig.~\ref{fig:densityPlot}. It displays the spin polarization as a function of frequency and dressing parameter $x$, with the calculated energy levels overlaid for comparison. This representation makes weak higher-order resonances easier to identify across a continuous range of dressing parameters. For each dressing parameter $x$, the baseline of the corresponding frequency scan was normalized to unity. The latter was achieved by averaging the spin polarization in the transition-free region at frequencies above \SI{2700}{\hertz}, ensuring consistent comparison across spectra.

\ifpreprint
    \section{Conclusion}
\fi

In conclusion, we have experimentally investigated the dressed states of proton spins for the first time. Using the quantum Rabi model, we calculated the expected energy-level structure and transition frequencies and observed the corresponding resonances in a Rabi-type spectroscopic measurement. Our results confirm the model's predictions in the high-photon-number regime and reveal higher-order transitions that had previously not been observed in nuclear systems. These measurements provide direct experimental access to dressed-state dynamics, an understanding that is essential for precision spin-manipulation experiments, including searches for electric dipole moments and advanced magnetic-resonance techniques. Moreover, controlled spin dressing offers a route to compensating systematic frequency shifts such as the Bloch--Siegert shift. These results establish nuclear spin dressing as an experimentally accessible platform for exploring dressed-state physics in two-level systems beyond the rotating-wave approximation.

\begin{acknowledgments}
We gratefully acknowledge the excellent technical support by S.~Bosco, L.~Meier, and V.~Vitacca from the University of Bern. We also thank the Albert Einstein Center for Fundamental Physics (AEC) of the University of Bern for the support through their visitor program. This work was supported through the European Research Council under the ERC Grant Agreement no. 715031 (BEAM-EDM) and the Swiss National Science Foundation under grants no. 163663, 181996, and 230596. 
\end{acknowledgments}

\section*{Conflict of Interest}

The authors have no conflict of interest to disclose.

\section*{Author Contributions}

Contribution of all authors according to ANSI/NISO~\cite{nisocreditworkinggroup_ansi_2022}. 

\textbf{Conceptualization:} I. Schulthess, P. Hautle, F. M. Piegsa; 
\textbf{Data curation:} I. Schulthess;
\textbf{Formal analysis:} I. Schulthess;
\textbf{Funding acquisition:} I. Schulthess, F. M. Piegsa;
\textbf{Investigation:} I. Schulthess, F. M. Piegsa;
\textbf{Methodology:} I. Schulthess, F. M. Piegsa;
\textbf{Project administration:} F. M. Piegsa;
\textbf{Resources:} P. Hautle, F. M. Piegsa;
\textbf{Software:} I. Schulthess;
\textbf{Supervision:} F. M. Piegsa;
\textbf{Validation:} I. Schulthess, F. M. Piegsa;
\textbf{Visualization:} I. Schulthess;
\textbf{Writing - original draft:} I. Schulthess;
\textbf{Writing - review \& editing:} I. Schulthess, A. Fratangelo, P. Hautle, P. Heil, G. Markaj, M. Persoz, C. Pistillo, J. Thorne, F. M. Piegsa.

\section*{Data Availability}
The data and analysis that support the findings of this study are openly available in a Github repository~\cite{schulthess_ivoschulthessprotonnmr_dressedstates_2023}.


\bibliography{ref.bib}

\end{document}